\documentclass[12pt]{article}
\usepackage{epsfig}
\begin{document}

\renewcommand{\thefootnote}{\fnsymbol{footnote}}

\baselineskip=20pt
\begin{center}
{\Large \bf Measurement of the Neutron Flux Produced by Cosmic-Ray Muons
with LVD at Gran Sasso}
\end{center}
\baselineskip=14pt

\begin{center}
\vspace{0.2cm}

{\large \bf LVD Collaboration}

\vspace{0.3cm}
M.Aglietta$^{14}$, E.D.Alyea$^{7}$, P.Antonioli$^{1}$,
G.Badino$^{14}$, G.Bari$^{1}$, M.Basile$^{1}$,
V.S.Berezinsky$^{9}$, F.Bersani$^{1}$, M.Bertaina$^{8}$,
R.Bertoni$^{14}$, G.Bruni$^{1}$, G.Cara Romeo$^{1}$,
C.Castagnoli$^{14}$, A.Castellina$^{14}$, A.Chiavassa$^{14}$,
J.A.Chinellato$^{3}$, L.Cifarelli$^{1,\dagger}$, F.Cindolo$^{1}$,
A.Contin$^{1}$, V.L.Dadykin$^{9}$,
L.G.Dos Santos$^{3}$, R.I.Enikeev$^{9}$, W.Fulgione$^{14}$,
P.Galeotti$^{14}$, P.Ghia$^{14}$, P.Giusti$^{1}$, F.Gomez$^{14}$,
R.Granella$^{14}$, F.Grianti$^{1}$, V.I.Gurentsov$^{9}$,
G.Iacobucci$^{1}$, N.Inoue$^{12}$,
E.Kemp$^{3}$, F.F.Khalchukov$^{9}$, E.V.Korolkova$^{9}$,
P.V.Korchaguin$^{9}$, V.B.Korchaguin$^{9}$, 
V.A.Kudryavtsev$^{9\dagger\dagger}$\footnote{e-mail: 
v.kudryavtsev@sheffield.ac.uk},
M.Luvisetto$^{1}$,
A.S.Malguin$^{9}$, T.Massam$^{1}$, N.Mengotti Silva$^{3}$,
C.Morello$^{14}$, R.Nania$^{1}$, G.Navarra$^{14}$,
L.Periale$^{14}$, A.Pesci$^{1}$, P.Picchi$^{14}$,
I.A.Pless$^{8}$, V.G.Ryasny$^{9}$,
O.G.Ryazhskaya$^{9}$, O.Saavedra$^{14}$, K.Saitoh$^{13}$,
G.Sartorelli$^{1}$,
M.Selvi$^{1}$, N.Taborgna$^{5}$, V.P.Talochkin$^{9}$,
G.C.Trinchero$^{14}$, S.Tsuji$^{10}$, A.Turtelli$^{3}$,
P.Vallania$^{14}$, S.Vernetto$^{14}$,
C.Vigorito$^{14}$, L.Votano$^{4}$, T.Wada$^{10}$,
R.Weinstein$^{6}$, M.Widgoff$^{2}$,
V.F.Yakushev$^{9}$, I.Yamamoto$^{11}$,
G.T.Zatsepin$^{9}$, A.Zichichi$^{1}$

\medskip

$^{1}$ {\it University of Bologna and INFN-Bologna, Italy}\\
$^{2}$ {\it Brown University, Providence, USA}\\
$^{3}$ {\it University of Campinas, Campinas, Brazil}\\
$^{4}$ {\it INFN-LNF, Frascati, Italy}\\
$^{5}$ {\it INFN-LNGS, Assergi, Italy}\\
$^{6}$ {\it University of Houston, Houston, USA}\\ 
$^{7}$ {\it Indiana University, Bloomington, USA}\\
$^{8}$ {\it Massachusetts Institute of Technology, Cambridge, USA}\\
$^{9}$ {\it Institute for Nuclear Research, Russian Academy of
Sciences, Moscow, Russia}\\
$^{10}$ {\it Okayama University, Okayama, Japan}\\ 
$^{11}$ {\it Okayama University of Science, Okayama, Japan}\\
$^{12}$ {\it Saitama University of Science, Saitama, Japan}\\
$^{13}$ {\it Ashikaga Institute of Technology, Ashikaga, Japan}\\
$^{14}$ {\it University of Torino and INFN-Torino, Italy} \\
\indent {\it Institute of Cosmo-Geophysics, CNR, Torino, Italy}\\
\vspace{0.2cm}
$^{\dagger}$ {\it now at University of Salerno and INFN-Salerno, Italy}\\
$^{\dagger\dagger}$ {\it now at University of Sheffield, UK}\\
\end{center}
\vspace{0.2cm}
\begin{center}
{\large \bf Abstract\\}
\end{center}

The flux of muon-produced neutrons far away from the muon track
may constitute a background for the underground detectors searching for
rare events. The muon events collected by the first LVD tower
from March, 1996, to February, 1998, (1.56 years of live time) were used
to estimate the neutron flux at various distances from the muon track
or muon-produced cascade.

\pagebreak

\section{Introduction}

It is well known that neutrons produced by cosmic-ray muons can move far
away from the muon tracks or muon-initiated cascades and contribute to 
the background
for large underground experiments searching for rare events such as
neutrino interactions, proton decay etc. (see, for example, 
Khalchukov et al., 1983). At present there are few measurements of the
muon-produced neutron flux at large depths underground 
(Bezrukov et al., 1973, Enikeev et al., 1987, Aglietta et al., 1989).
In these experiments the detection of both muon and neutron was required
but the distance between them was not measured.
In this work we analyse the most general case, when both
muon and neutron are detected by LVD (Large Volume Detector at the
underground Gran Sasso Laboratory) and the distance between them (or
between neutron and muon-initiated cascade) is known.
The present experiment is carried
out with the same scintillator ($C_n H_{2n}$, $<n>\approx 9.6$)
as used in aforementioned earlier experiments.

\section{Detector and Data Analysis}

The data presented here were
collected with the 1st LVD tower during 13639 hours of live time.
The 1st LVD tower contains 38 identical modules.
Each module con\-sists of 8 scintillation counters, each 1.5 m $\times$
1.0 m $\times$ 1.0 m, and 4 layers of limited
streamer tubes (tracking detector) attached to the bottom and to one
vertical side of the metallic supporting structure. 
Each counter is viewed by 3 photomultiplier tubes (PMT) on top of
the counter. A detailed description of the
detector was given in Aglietta et al. (1992). 
The depth of the LVD site averaged over the muon flux is about 3650 hg/cm$^2$
which corresponds to mean muon energy underground of about 270 GeV.
Each scintillation counter is self triggered by the three-fold coincidence
of the PMT signals after discrimination. The high-energy threshold (HET)
is set at 4-5 MeV for inner counters.
During the 1 ms time period following an HET trigger, a low-energy 
threshold (LET)
is enabled for counters belonging to the same quarter of the tower
which allows the detection of the 2.2 MeV photons from neutron
capture by protons. 
Further on we will
consider only the signals induced by neutrons in the inner counters, where
the LET is low enough (0.8 MeV) to allow high neutron detection efficiency,
while the background rate is quite small. 
138 counters were considered as inner ones. 

All muon events were divided into two classes: i) 'muons' -- single 
muon events, where
a single muon track is reconstructed (small cascades cannot be excluded), 
and ii) 'cascades' -- there is no clear single muon track but the 
energy release is high
enough to indicate that at least one muon is present; 
such events may be due to either muon-induced cascades or multiple muons.

Each neutron ideally should generate
two pulses:
the first pulse above the HET is due to the recoil protons from $n-p$ elastic 
scattering (its amplitude is proportional to and
even close to the neutron energy);
the second pulse, above the LET in the time gate of about 1 ms is due to the
2.2 MeV gamma from neutron capture by a proton. The sequence of two pulses
(one above the HET and one above the LET) was the signature of neutron 
detection.
The energy of the first pulse (above HET) was measured and attributed
to the neutron energy. Note that really this is not a neutron energy
but the energy transferred to protons in the scintillator and measured by 
the counter. The
distance between the counter in which a neutron was detected, and the muon
track was calculated. For single muons this was the minimal distance 
between the center of the counter
which detected a neutron and the centers of counters traversed by the muon. 
The precision (about 1 m) is really
restricted by the fact that neither the point of neutron production nor the
point of neutron capture are known with better accuracy.
If the neutron is detected in a counter crossed by the muon (distance is less
than 1 m) the energy cannot be attributed to the neutron alone but 
includes the muon energy loss. For cascade events 
we calculated the minimal
distance between the center of counter where the neutron was detected and the
centers of the counters struck by the cascade excluding that with the neutron.

Pairs of seemingly time-correlated pulses might also be produced by 
random coincidences
of high-energy pulses (above HET) due to muon or cascade energy release
and low-energy pulses (above LET) due to local radioactivity. The
counting rate of such random coincidences per muon event 
is determined by the counting rate of background
pulses in the time gate. To evaluate this background the counting rate
of low-energy pulses in the counters in the absence of high-energy pulses 
was measured.
The true neutron flux per unit energy and unit distance was calculated as a 
difference between the total number of correlated pairs observed 
and the number of pairs expected due to random coincidences.

\section{Results and Discussion}

\begin{figure}[htb]
\begin{center}
\epsfig{file=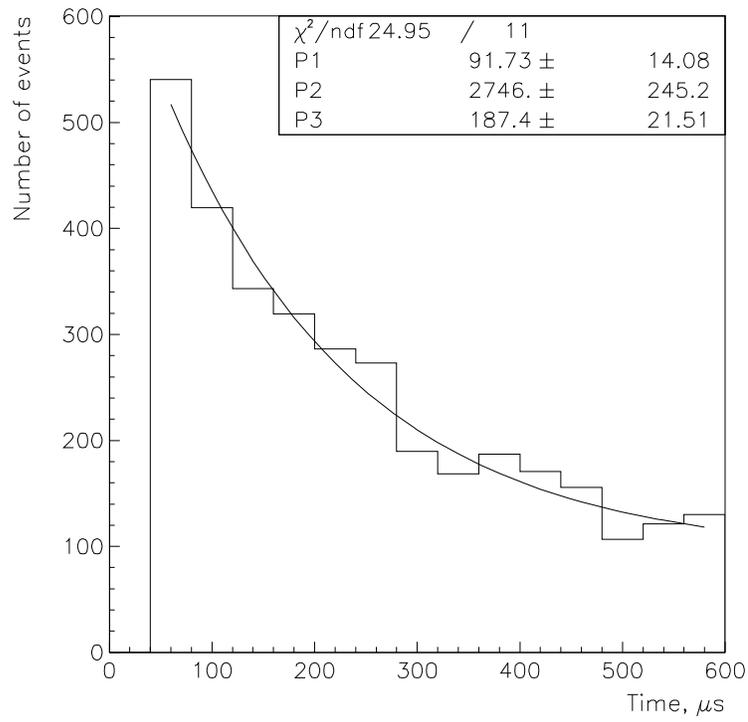,height=12cm}
\caption {Time distribution of pulses of 2.2 MeV gammas from
neutron capture by protons.}
\label{fig1}
\end{center}
\end{figure}

\begin{figure}[htb]
\begin{center}
\epsfig{file=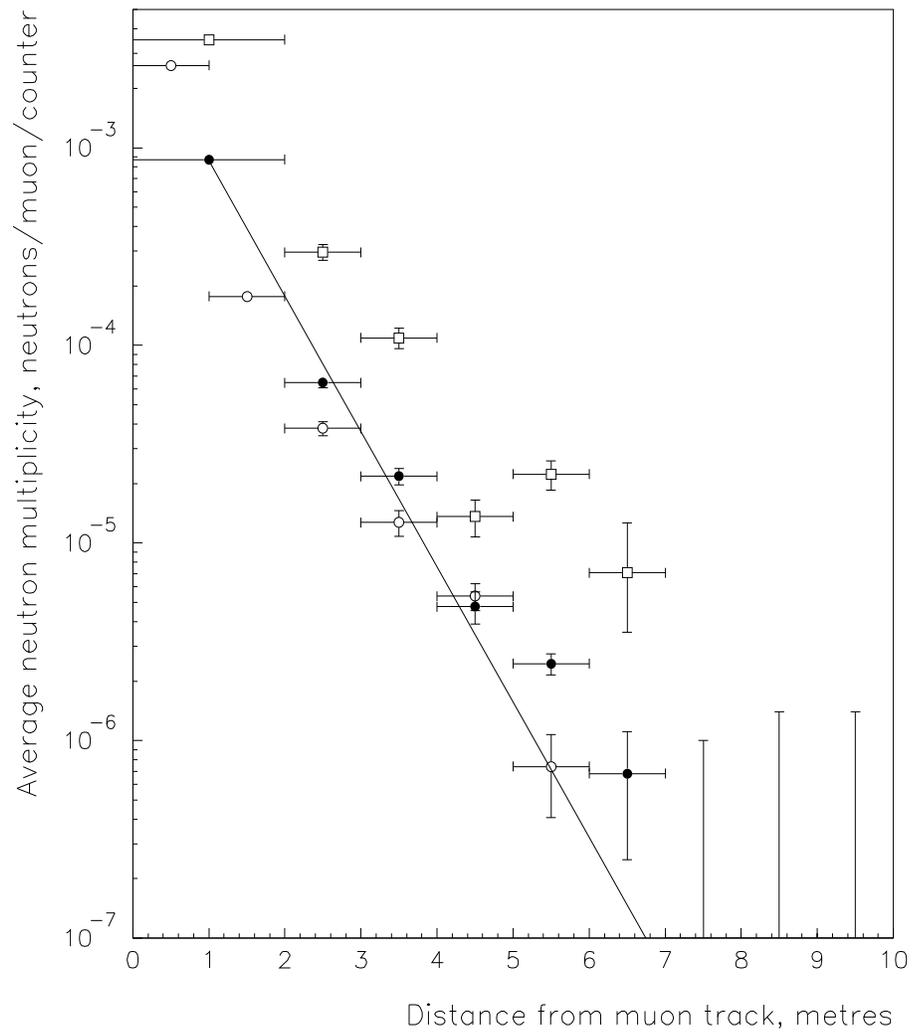,width=12cm,height=15cm}
\caption {Neutron multiplicity versus distance from the muon
track or cascade core. {\it Open circles} -- single muons;
{\it open squares} -- cascades; {\it filled circles} and 
{\it solid curve} -- all
events and exponential fit (upper limits are shown
at $R >$ 7 m.} \label{fig8}
\end{center}
\end{figure}

A typical time distribution of the LET pulses after an HET pulse is shown
in Figure \ref{fig1}. The plot includes
single muon events with all HET energies and distances (1--2) m
from the muon track.
Although the expected background has already been
subtracted, as described in Section 2, 
the figure shows an exponential superimposed on a flat distribution
of background pulses. This implies that the real level of background
of LET pulses is higher in the counters with HET pulses than without HET pulses.
The measured distribution was fitted with the
following formula:
$dN/dT = B + N_n/\tau \cdot exp(-t/\tau)$,
where $B=92^{+13}_{-15}$ is the constant term (residual background)
per bin, $N_n=2746^{+273}_{-218}$ is the total number of neutrons, 
and $\tau=187^{+24}_{-19}$ $\mu$s
is the mean time of neutron capture. The value of $\tau$ is in good
agreement with previous measurements 
(Aglietta et al., 1989, and references therein). To obtain the numbers
of neutrons at various distances from the muon track, similar distributions
were fitted to the above equation with fixed value of 
$\tau=190$ $\mu$s.
The average neutron multiplicity per event (whether single muon or cascade)
per counter
is plotted against distance from the muon track or cascade core 
in Figure \ref{fig8} (single muons - open circles,
cascades - open squares). Only the statistical errors from the fits
are shown. Horizontal bars show the range of distances for each point.
The total contributions to the neutron flux of single muons and 
cascades are found to be roughly comparable.
As the number of reconstructed single muon events exceeds that of 'cascade'
events by an order of magnitude, the number of neutrons per cascade is
several times higher than the number of neutrons per single muon event.
This supports the results of previous measurements 
(Bezrukov et al., 1973, Enikeev et al., 1987,
Aglietta et al., 1989) and early estimations
(Ryazhskaya \& Zatsepin, 1966).
The results of combined treatment of single muons and cascades
are shown by filled circles. Only upper limits to the neutron multiplicity
can be obtained for last three bins.
The exponential fit to the all-event distribution is shown by the solid curve:
$F = A \cdot exp(-R/<R>)$
where $A=(4.17 \pm 0.17) \cdot 10^{-3}$ neutrons/(muon event)/counter
and $<R>=(0.634 \pm 0.012)$ m.

Note that the LVD is not a uniform detector, and there are several
ten-centimeter air gaps between modules and several centimeter gaps
between the counters in a module. This means that the neutrons, as well as
other secondary particles, can escape from the counters where they were
produced and reach another counter (possibly quite far from the original one)
by way of low density air gaps.  

\begin{figure}[htb]
\begin{center}
\epsfig{file=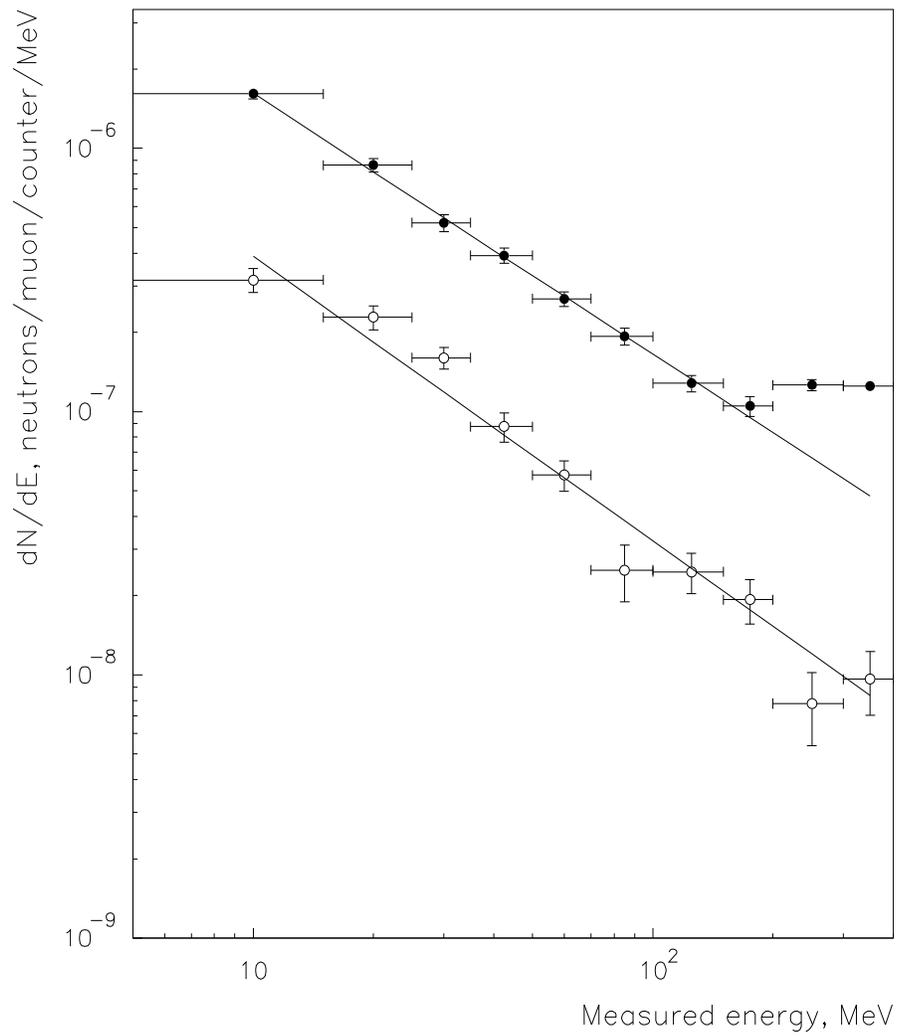,width=12cm,height=15cm}
\caption {Neutron flux versus energy of HET pulses at $R >$ 1 m
({\it filled circles}) and $R >$ 2 m ({\it open circles}).} \label{fig9}
\end{center}
\end{figure}

The energy of a trigger pulse in a counter 
can be attributed to the
neutron kinetic energy if: 1) there is no energy loss of the muon nor that of
secondary particles (other than neutrons) in this particular counter;
2) all neutron kinetic energy is tranferred to protons inside this
counter; 
3) the energy deposited by a recoil proton is proportional to the 
pulse amplitude
and this proportionality is the same as for electron pulses.
Although these conditions are not strictly satisfied,
we assume (at zero approximation) that the measured spectrum of 
HET pulses at large enough
distances from the muon track corresponds to the neutron energy spectrum
(near the point of neutron capture, within a sphere
with a diameter of about 1 m).
Such a spectrum is presented in Figure \ref{fig9} by filled circles for
distances $R >$ 1 m. Only statistical errors are shown.
Horizontal bars show the range of energies for each point.
To check the contamination of the distribution
by energy deposited by secondary particles (other than neutrons)
we plotted also the energy spectrum of HET pulses at distances
$R >$ 2 m (open circles). It is obvious that the contributions of
secondary particles of all kinds should decrease with $R$. 
Both data samples show
similar behaviour at $E <$ 200 MeV. At $E >$ 200 MeV the points for
$R >$ 1 m (filled circles) are higher than is expected from
the general trend of the spectrum. This can be explained by 
contamination from the energy loss of muons (this is a region near the peak of
muon energy release in the counter) and cascade particles. 
There is no such excess of events at 200-400 MeV at distances
$R >$ 2 m. This means that the contribution of secondary particles
other than neutrons to the neutron spectrum at high energies is negligible.
Both spectra were fitted with power-law functions with two free parameters:
$dN/dE=A \cdot E^{-\alpha}$.
The energy bins 200-300 MeV and 300-400 MeV were excluded from
the analysis of data at $R >$ 1 m. The results of the fits are:
$A=(1.58 \pm 0.14) \cdot 10^{-5}$ neutrons/(muon event)/counter/MeV, 
$\alpha=0.99 \pm 0.02$ for $R >$ 1 m,
and $A=(4.67 \pm 0.71) \cdot 10^{-6}$ neutrons/(muon event)/counter/MeV, 
$\alpha=1.08 \pm 0.04$ for $R >$ 2 m.
The errors are statistical only. 
The slopes of the two spectra are in good agreement, a more quantitative 
indication that the contributions of energy losses of particles 
other than neutrons are not
very important, if not negligible at energies less than 200 MeV or distances 
more than 2 meters. The slope of the spectrum is also in
reasonable agreement with the results of Monte Carlo simulations
(Dementyev et al., 1997).
The units used in Figure \ref{fig8} and \ref{fig9} can be converted 
to more convenient
ones (m$^{-2}$) assuming that each counter has an average area of about 
1.5 m$^{2}$ orthogonal to the direction of neutron flux and dividing 
each value by
the neutron detection efficiency (about 0.6 for MeV-neutrons uniformly
distributed in the counter volume).

Finally, we calculated the average number of neutrons produced by a muon per
unit path length in liquid scintillator using the formula:
$<N> = N_n \cdot Q /(N_c \cdot L \cdot \epsilon)$,
where $<N>$ is the average number of neutrons produced by a muon
per 1 g/cm$^2$ of its path in scintillator, $N_n$ is the total
number of neutrons at all distances from the track
(result of the fit similar to that shown in Figure \ref{fig1}),
$N_c$ is the number of counters crossed by muons, $L$ is the mean path
length of a muon inside the counter, $\epsilon$ is the efficiency of
neutron detection in the inner LVD counters, and $Q$ is the correction factor
which takes into account the neutron production in iron of the supporting
structure and counter walls.
The fit of the all-data
sample gives $N_n=(2.34 \pm 0.46) \cdot 10^4$ neutrons. The error 
includes both statistical and systematic uncertainties.
The total number of counters 
crossed by all muons can be calculated directly only for single muon events 
when the muon track is well reconstructed. The estimation for all events
results in the value of
$N_c=(3.3 \pm 0.2) \cdot 10^6$.  
The mean path length of muons in the scintillation counter is equal to
$65 \pm 7$ g/cm$^2$ (Aglietta et al., 1989).
The efficiency of neutron detection by one scintillation counter has been
measured with a source and calculated by Monte Carlo techniques 
(see, for example,
Aglietta et al., 1989, 1992). For MeV-neutron sources uniformly 
distributed in the
counter volume the efficiency of neutron detection is $0.6 \pm 0.1$.
The correction factor $Q$ takes into account the neutron
production in iron which should be subtracted since we want to calculate
that in scintillator only. It was obtained by Aglietta et al. (1989) for LSD,
$Q=0.61 \pm 0.04$. Similar estimation for LVD gives $Q=0.85 \pm 0.10$.
Finally, one gets $<N>=(1.5 \pm 0.4) \cdot 10^{-4}$
neutrons/(muon event)/(g/cm$^2$).
The value of $<N>$ is
3.5 times smaller than that obtained with LSD detector at larger depth.
This difference cannot be easily explained even by large systematic
uncertainties and the difference of the depths of the detector sites.

\section{Conclusions}

The muon events collected by the first LVD tower 
(1.56 years of live time) were used
to estimate the neutron flux at various distances from the muon track
(or from the muon-produced cascade). The neutron flux decreases by
more than three orders of magnitude at distances more than 5 meters
from muon track or cascade core. The average number of neutrons
produced per muon per g/cm$^2$ of its path in the liquid
scintillator is found to be $(1.5 \pm 0.4) \cdot 10^{-4}$ 
neutrons/(muon event)/(g/cm$^2$). Under the assumptions mentioned in
the previous section the neutron differential energy spectrum 
in the kinetic energy range (5 -- 400) MeV was found to
follow a power law with exponent close to -1.

\section{Acknowledgements}

We wish to thank the staff of the Gran Sasso Laboratory
for their aid and collaboration. This work is supported by the
Italian Institute for Nuclear Phy\-sics (INFN) and in part by the
Italian Ministry of University and Scientific-Technological
Research (MURST), the Russian Ministry of Science and Technologies,
the US Department of Energy, the US National
Science Foundation, the State of Texas under its TATRP program,
and Brown University.
One of us (V. A. Kudryavtsev) is grateful to the University of 
Sheffield for hospitality.

\vspace{1cm}
\begin{center}
{\Large\bf References}
\end{center}
Aglietta, M. et al. 1989, Nuovo Cimento 12C, 467.\\
Aglietta, M. et al. 1992, Nuovo Cimento 105A, 1793.\\
Bezrukov, L. B. et al. 1973, Sov. J. Nucl. Phys. 17, 987.\\
Dementyev, A. et al. 1997, Preprint INFN/AE-97/50.\\
Enikeev, R. I. et al. 1987, Sov. J. Nucl. Phys. 46, 1492.\\
Khalchukov, F. F. et al. 1983, Nuovo Cimento 6C, 320.\\
Ryazhskaya, O. G. \& Zatsepin, G. T. 1966, Proc. 9th Intern. Cosmic Ray 
Conf. (London) 3, 987.\\
Zatsepin, G. T. et al. 1989, Sov. J. Nucl. Phys. 49, 266.\\

\pagebreak

\pagebreak

\pagebreak

\end{document}